# Carbon Nanotube Wools
# Directly from $CO_2$ By Molten Electrolysis:
# Value Driven Pathways to Carbon Dioxide Greenhouse Gas Mitigation

Marcus Johnson[1], Jiawen Ren[1], Matthew Lefler[1], Gad Licht[1], Juan Vicini[1], Stuart Licht[1],*

[1]Dept. of Chemistry, George Washington University, Washington DC 20052.

*Correspondence: slicht@gwu.edu (S.L.).

**Summary / Context & Scale**
A climate mitigation comprehensive solution is presented through the first high yield, low energy synthesis of macroscopic length carbon nanotubes ("CNT") wool from $CO_2$ by molten carbonate electrolysis, suitable for weaving into carbon composites and textiles. Growing $CO_2$ concentrations, the concurrent climate change and species extinction can be addressed if $CO_2$ becomes a sought resource rather than a greenhouse pollutant. Inexpensive carbon composites formed from carbon wool as a lighter metal, textiles and cement replacement comprise a major market sink to compactly store transformed anthropogenic $CO_2$. 100x-longer CNTs grow on Monel versus steel. Monel, electrolyte equilibration, and a mixed metal nucleation facilitate the synthesis. $CO_2$, the sole reactant in this transformation, is directly extractable from dilute (atmospheric) or concentrated sources, and is cost constrained only by the (low) cost of electricity. Today's $100K per ton CNT valuation incentivizes $CO_2$ removal.

**In Brief**

The first electrosynthesis of carbon nanotube wool is shown, and the only reactant, $CO_2$, becomes a useful, valuable resource rather than a greenhouse pollutant as a comprehensive response to removal of anthropogenic carbon dioxide.

**Introduction**

The planet is heating up. The atmospheric $CO_2$ concentration, which had cycled at ~235 $\pm$ ~50 ppm for 400,000 years until 1850 is currently at 406 ppm and rising (NASA: Global Climate Change, 2017; NASA: Global Climate Change, March 2017). In l824 Fourier found our atmosphere insulates the Earth; in 1864 Tyndall measured $CO_2$ infrared absorption heating up the Earth. The thermal interactive dynamics of the land/sea/air are complex, but by 1896 Nobel laureate Arrhenius (physicist, chemist, and electrolytic theory) estimated the greenhouse effect magnitude and wrote in 1908 in the Worlds in the Making: "any doubling of the percentage of carbon dioxide in the air would raise the temperature of the earth's surface by 4°C; ... The enormous combustion of coal by our industrial establishments suffices to increase the percentage of carbon dioxide in the air to a perceptible degree". In the past century atmospheric $CO_2$ rose by one third, and during this time the average land and ocean temperature increased by 1.2°C to 12.1°C (in early 2017). Among global warming effects is the extinction of a substantial fraction of the planet's species (Urban, 2015), and the incidence of climate disruptions (on biodiversity, drought, major storms, etc.) is on the rise (Pimm, 2009). Through 2008, $CO_2$ was regarded as such a stable molecule that its transformation into a non-greenhouse gas posed a major challenge (Prakash, et al., 2008). In 2009, we presented a solar theory that thermal energy, such as sub-band gap energy, unused by solar cells could greatly improve the efficiency of $CO_2$ splitting (Licht, 2009). In 2010 we demonstrated that an efficient (37%) concentrator photovoltaic cell, could become even more solar efficient by directing the unused solar thermal component of sunlight to heat and split $CO_2$ in a molten carbonate electrolyte. Both carbon monoxide solar fuel (at 950°C) and solid carbon (at 750°C) products were separately demonstrated (Licht, et al., 2010). Subsequent to this, alternative high temperature redox chemistries were developed to apply this solar thermal electrochemical process (STEP) to the generation of a number of staple products including ammonia, organics, calcium oxide, iron and methane fuel (Li, et al., 2011; Li, et al., 2015; Li, et al., 2016; Licht, 2011; Licht, et al., 2012; Licht, et al., 2014; Wu, et al., 2016; Zhu, et al., 2016). However, fuels from sunlight including methane, syngas, or a coal equivalent (Li, et al., 2011; Li, et al., 2015; Wu, et al., 2016), are in the value range of $100 per ton product providing no financial incentive for $CO_2$ mitigation. Furthermore, use of the fuel releases the captured $CO_2$ back into the atmosphere.

     The first facile high yield, low energy synthesis of <u>macroscopic length</u> carbon nanotubes (CNTs) from $CO_2$ by molten carbonate electrolysis is demonstrated here. This CNT "wool" is of length and uniformity suitable for weaving into carbon composites and textiles. Such products have a contemporary market value in the $100K per ton range (Cheaptubes, 2017). Monel cathode substrates, electrolyte equilibration, and a mixed metal (NiChrome) nucleation facilitate the synthesis. $CO_2$ is the sole reactant in this CNT transformation. The process is constrained by the (low) cost of electricity. In 2015, we presented the high yield, low energy electrolytic splitting of $CO_2$ to carbon nanofibers and nanotubes from the molten carbonate electrolysis, (Ren, et al., 2015) and demonstrated that the less expensive natural carbon isotope mix ($^{12}C_{0.99}^{13}C_{0.01}O_2$, rather than $^{13}CO_2$) produced the more expensive (carbon nanotube, rather than nanofiber) product, (Ren and Licht, 2016). We have termed this high yield, molten carbonate electrolytic transformation of $CO_2$ to CNTs as the C2CNT process. However, the complex synthesis involved (i) a zinc coated steel cathode, (ii) a pre-CNT low current activation process to initiate the cathode nucleation, and then generated useful, but only shorter (< 100 μm length) CNTs. High yield, low electrolysis voltage, and high uniformity CNTs, as produced from $CO_2$ dissolved in a variety of alkali and alkali-earth carbonate electrolytes, were demonstrated (Ren, et al., 2017; Wu, 2016), and high conductivity CNTs, such as boron and other doped CNTs are readily formed by the controlled addition of impurities to the electrolyte during the electrosynthesis as exemplified in the Supplemental Information. Due to the expense, energy intensity and complexity of the conventional chemical vapor deposition (CVD) synthesis and direct spin methodologies, industrial CNTs currently are valued in the $100K ($85-$450K) per ton range (Cheaptubes, 2017; Janas and Koziol, 2016; Magrez, et al., 2010; Jia and Wei, 2017) and do not use $CO_2$ as a reactant. In C2CNT, when hot flue gas is provided as a reactant, a hot $CO_2$ source for CNTs is provided and the greenhouse gas transforms to a useful product. The electrolysis, constrained by the 4 e- reduction consumes as little as $50 of electricity per ton of $CO_2$ when the oxy-fuel energy improvement is taken account (the oxy-fuel improvement uses the

$O_2$ by-product of the CNT production to improve the electrical generation efficiency) (Lau, et al., 2016; Licht, 2017).

**Results and Discussion**

Figure 1 illustrates the new C2CNT methodology advances compared to prior C2CNT molten carbonate electrosyntheses. The first CNT wool from $CO_2$ by molten carbonate electrolysis is demonstrated, suitable for weaving into carbon composites and textiles. 100x longer CNTs grow on Monel versus copper. Monel cathodes, electrolyte equilibration, and a mixed metal nucleation facilitate the synthesis. The previous synthetic pathway led to only short CNTs. Longer electrolyses had produced only thicker and more convoluted, but not longer, CNTs (Wu, 2016). That methodology was dependent on a zinc coated steel cathode, a pure Ni anode, and a low current pre-electrolysis activation step (Ren, et al., 2015; Ren, et al., 2017; Ren and Licht, 2016; Wu, 2016). As delineated in the Supplemental Information, an intermediate, C2CNT electrolysis explored here, removes the requirement of a zinc coating leading to the exploration of a variety of new electrolysis bare cathode substrates, but requires the addition of Ni metal powder directly to the molten carbonate electrolyte. On the figure right side, the optimized C2CNT methodology is more straightforward and produces a high yield of macroscopic length CNT wool. Monel cathodes and Nichrome anodes are effective, and subsequent to molten electrolyte equilibration for 24 hours, the C2CNT electrolysis is conducted directly without pre-electrolysis activation steps. We have previously measured that molten lithium carbonate requires several hours to achieve an equilibrium concentration of 0.29 $\pm$0.04 m $Li_2O$ at 750°C in accord with step (Ren, et al., 2015): $Li_2CO_3 \rightleftharpoons Li_2O + CO_2$. Sufficient electrolyte equilibration time is observed to be a significant determining factor in C2CNT growth of uniform CNT wool. Oxide/peroxide/superoxide speciation in molten carbonates is complex, time and cation dependent, and affects electrochemical charge transfer in this media. For example between 727°C to 827°C, the relative proportions of oxide, peroxide and superoxide in Li, Li/Na. Li/K & Li/Na/K carbonate exhibit no superoxide (<0.01%), while Na/K carbonate contains ~45% superoxide. While the oxide dominates in the pure Li electrolyte (with 99.97% oxide relative to peroxide), this proportional is 98%, 96%, 95% and only 10% respectively in Li/Na. Li/K & Li/Na/K and Na/K carbonates (Cassir, et al., 1993).

Figure 2 presents the observed growth mechanism of the macroscopic CNT wool introduced in this study. A molten $Li_2CO_3$ is aged one day at 770°C, and then a Nichrome anode and a Monel cathode immersed, and at 770°C electrolysis is conducted at constant current (in this case 0.1 A cm$^{-2}$). Initially a thin carbon (as confirmed by EDS) coating forms on the Monel cathode substrate surface. The carbon coating is thin as confirmed by the relative transparency to e- beam in the SEM. We have observed that this layer is even thinner when grown on pure copper (not shown). We suggest that this opens a pathway to the ready electrochemical synthesis of graphene sheets in molten carbonates (the topic of a later study). Complete coverage of the substrate by this graphene coating under continuous applied electrolysis current is self-terminating, and carbon growth continues as carbon nano-particle growth. This component of the growth is similar to the previously observed Stranski−Krastanov thin film growth, and Volmer−Weber island growth CVD growth regimes (Eadlesham, et al., 1990; Yan, et al., 2003) that control the stacking order in graphene bilayers (Ta, et al., 2016).

The morphology of carbon nano-particle growth is highly dependent on the initial metal substrate and the transition metals available to act as nucleation growth points. It is observed (by EDS analysis of the elemental composition at localized points on the graphene coating) that transition metals form nucleation points, which initiate growth of the observed grown CNTs. Here, the source of the transition metals is the low-level oxidative release from the anode. For example, nickel, $Ni^{2+}$ is soluble, but only at ppm concentrations in these molten carbonate electrolytes (Lee, et al., 1996), and the anodic release of $Ni^{2+}$ slows as a stabilized nickel oxide layers forms on the anode surface (Licht and Wu, 2011). Both nickel and chromium are available when NiChrome, rather than pure nickel (Ni 200) is used as the anode. This availability of nickel and chromium as co-nucleation metals results in the most uniform, consistent and longest CNTs we have electrosythesized to date.

Figure 3 present the high quality CNT wool product consistently obtained in replicate syntheses subsequent to cathode extraction and washing after an 18-hours electrolysis at 0.1 A $cm^{-2}$ under these electrolysis conditions (Monel cathode, Nichrome anode, 24 hour equilibrated 770°C $Li_2CO_3$). The carbon product is immediately observed to be different then that from our prior molten carbonate electrosyntheses. This product is wool-like and fluffy, rather than the powdery or sandy. In the figure, this difference is evidenced by the macroscopic length (100 fold longer) than the prior 5 to 50 µm length of previous syntheses. As with previous, optimized C2CNT electrosyntheses, the typical C2CNT coulombic yield (in which 100% is equivalent to a 4 e- conversion of all applied charge reducing the tetravalent carbon) is > 90% (and higher with careful product recovery). Additionally, in the case of the new, synthesis the carbon product consists of ~95% macroscopic CNTs. As seen in figure, the CNT product ranges 0.4 to 1.2 mm long although optical (not shown), rather than SEM microscopy of the same product shows several outlier CNTs that are over 2 mm long. As seen in the lower portion of Figure 4, the inter-graphene separation of the CNTs' walls by TEM exhibits the typical 0.342 nm variation (Ren and Licht, 2016), and the CNTs' diameters ranges from 0.5 to 1.5 µm. The inner diameter of the tubes varies by as much as a factor of three. As we have recently demonstrated, we will be able to refine and further minimize the range of these parameters by careful control of the electrolyte composition, temperature and current density (Ren *et al*, 2017).

**Discussion**

A synthesis is introduced that forms the first electrolytic CNT wool and consists of a hundred-fold increase in the length of molten carbonate electrolyte CNTs. 100x-longer CNTs grow on Monel versus steel. Monel, electrolyte equilibration, and a mixed metal nucleation facilitate the synthesis. $CO_2$, the sole reactant in this transformation, is directly extractable from dilute (atmospheric) or concentrated sources, and is cost constrained only by the (low) cost of electricity. The Supplemental Information details the widely varying carbon morphologies produced at different cathode substrates during molten carbonate electrolysis, the ready addition of heteroatoms during the electrolysis, to achieve CNTs with difference properties, intermediate stages of the C2CNT synthesis methodology advance and the scalability of the C2CNT process. The low electrical cost, ease of synthesis and wide range of uniform CNTs accessible by C2CNT provide an economic incentive to treat $CO_2$ as a resource, rather than a pollutant, to encourage removal of this greenhouse gas to mitigate climate change.

Efficacious climate mitigation by $CO_2$ transformation requires a massive market, and product stability and compactness. The most compact form of captured carbon is through its transformation to solid carbon. CNTs are among the highest strength and most stable materials. CNT cost reduction by C2CNT, provides a preferred (lower mass per unit strength) to the mass metal market, and the CNT wool introduced here accelerates CNT demand as a building industry and textile material. Together these principal societal staples, when produced from $CO_2$, comprise an ample demand to markedly decrease atmospheric carbon. Scalability of the C2CNT process is delineated in the Supplemental Information. Initial scaling is efficiently applied to available concentrated, hot sources of $CO_2$, such as eliminating the $CO_2$ emission from industrial smoke stacks and simultaneously forming valuable CNT wool. Larger scale C2CNT can be achieved through direct elimination of atmospheric $CO_2$ using solar heat and solar to electric PVs. We have calculated that a surface area equivalent to less than 10% of the Sahara Desert is sufficient to remove all excess anthropogenic $CO_2$ by carbonate electrolysis in ten years, but open ocean areas may provide a more available surface. Heat exchange, between pretreated and $CO_2$ extracted air will be formidable; this may be facilitated with saltwater separation of salts and fresh water, as schematically represented in Figure 5.

**Experimental Procedures**

Electrolyses are driven galvanostatically in 50 g of 770°C molten lithium carbonate (99% Alfa Aesar). The electrolysis is contained in a pure alumina crucible (AdValue 99.6%). A variety of metals are explored as alternative cathodes and coiled as spiraled wires to form a horizontal flat disc of area 5 cm$^2$. Above this is a comparable, parallel anode as 5 cm$^2$ coiled wire of either pure nickel (Ni 200) or NiChrome wire. $CO_2$ pre-concentration is not required. During electrolysis, a carbon product accumulates at the cathode and oxygen evolves at the anode in accord with a $CO_2$ splitting reaction. We have previously utilized $^{13}$C isotope $CO_2$ to track and demonstrate that $CO_2$ originating from the gas phase serves as the renewable carbon building blocks in the observed CNT product (Ren and Licht, 2016) and the net reaction is in accord with:

Dissolution: $CO_2(gas) + Li_2O(soluble) \rightarrow Li_2CO_3(molten)$ (1)
Electrolysis: $Li_2CO_3(molten) \rightarrow C(CNT) + Li_2O(soluble) + O_2(gas)$ (2)
Net: $CO_2(gas) \rightarrow C(CNT) + O_2(gas)$ (3)

Subsequent to electrolysis the product remains on the C2CNT cathode, but falls off with congealed electrolyte when the cathode is extracted, cooled, and uncoiled, or simply pealed off from a planar Monel cathode. A molten process that helps avoid electrolyte congealing on the cathode is to briefly reverse the polarity of the electrodes at the electrolysis end during the extraction. This tends to bubble gas at the CNT containing electrode and discourage electrolyte from adhering during the extraction. Several aqueous washing routes are equally viable to separate any congealed (water soluble) electrolyte from the insoluble CNTs. Whereas lithium formate and lithium chloride have a high aqueous solubility at 20°C, the solubility is relatively low of lithium carbonate (respectively 39.3g, 83.5g and 1.3g per 100 g H$_2$O). Hence, the product is washed with copious deionized water, or more quickly washed with small amounts of HCl (forming lithium chloride) or formic acid (forming lithium formate) to dissolve electrolyte that had congealed with the product, and the product is dried. The respective solubilities of lithium chloride and formate are even higher at 100°C (138g and 128g per 100 g H$_2$O respectively), and this higher washing temperature does not affect the stable CNT product. Alternatively, the lithium carbonate electrolyte is removed via its higher solubility in aqueous ammonium sulfate solutions, or as lithium bicarbonate (LiHCO$_3$, formed as aqueous Li$_2$CO$_3$ under higher CO$_2$ partial pressure than in air, for example Li$_2$CO$_3$ under 1 bar of CO$_2$ has a 20°C solubility of 4.6g per 100 g $H_2O$). In this latter case, the washing solution with the CNT product is cycled and removed with higher and then lower pressure, which respectively dissolves, then any $Li_2CO_3$ separates by precipitation from the CNT product.

The washed carbon product is analyzed by PHENOM Pro-X Energy Dispersive Spectroscopy (EDS) on the PHENOM Pro-X SEM or FEI Teneo LV SEM and by TEM with a JEM 2100 LaB6 TEM. Raman spectroscopy was measured with a LabRAM HR800 Raman microscope (HORIBA) using 532.14nm wavelength incident laser light with a high resolution of 0.6 cm$^{-1}$. The synthesis yield, 100%x $C_{experimental}$ / $C_{theoretical}$, is determined by the measured mass of washed carbon product removed from the cathode, $C_{experimental}$, and the theoretical mass, $C_{theoretical}$ = (Q/nF)x(12.01 g C mol$^{-1}$) which is determined from Q, the time integrated charged passed during the electrolysis, F, the Faraday (96485 As mol$^{-1}$ e$^{-}$), and the n = 4 e- mol$^{-1}$ reduction of tetravalent carbon. Further details are presented in sections I and II of the Supplemental Information.

**Supplemental Information**

- I. **C2CNT electrosynthesis with bare (uncoated) cathodes and without pre-electrolysis low current activation.**
- II. **C2CNT Intermediate length CNTs with intermediate integrated electrolysis charge transfer.**
- III. **Improved admixing of sulfur, nitrogen and phosphorous (in addition to boron) to carbon nanotubes.**
- IV. **Scalability of the C2CNT process.**

**Acknowledgments**

We are grateful to the Unites States National Science Foundation grant 1505830 for partial support of this study.**References and Notes**

Cassir, M., Moutiers, G., Devynck, J., (1993). Stability and Characterization of Oxygen Species in Alkali Molten Carbonate: A Thermodynamic and Electrochemical Approach. J. Electrochem. Soc. 140, 3114-3123.

Cheaptubes.com. (2017). Cheaptubes: Industrial Grade Carbon Nanotubes. Available from: http https://www.cheaptubes.com/product-category/industrial-carbon-nanotubes-products/

Eadlesham, D. J., and Cerullo, M. (1990). Dislocation-Free Stranski-Krastow Growth of Ge on Si(100). Phys. Rev. Lett 64, 1943-1946.6

**Figure Legends**

**Figure 1.** Schematic representation of new synergistic pathways to form a high yield of macroscopic length CNT "wool" by electrolysis in molten carbonate. Middle: Prior C2CNT syntheses were dependent on a zinc coated steel cathode, a pure Ni anode, and a low current pre-electrolysis activation step. An intermediate, new C2CNT electrolysis removes the requirement of a zinc coating leading to the exploration of a variety of new cathode substrates. Right side: The optimized C2CNT pathway utilizes Monel cathodes and Nichrome anodes, molten electrolyte equilibration for 24 hours, and the electrolysis is conducted directly without pre-electrolysis activation steps. This pathway produces a high yield of macroscopic length CNT wool. Left side: experimental cell configuration used in these C2CNT experiments.

**Figure 2.** SEM of product formed in the initial electrolysis stages at the Monel cathode/equilibrated electrolyte interface.

**Figure 3.** SEM of the CNT wool product produced at the cathode from $CO_2$ during replicate syntheses of 770°C $Li_2CO_3$ electrolysis. Electrolysis is for 18 hours at 0.1 $Acm^{-2}$ (1.8 $Ah\ cm^{-2}$) between a NiChrome anode and a Monel cathode.

**Figure 4.** TEM presenting the range of these C2CNT generated CNT diameters. The lower left 10 nm scale TEM presenting the inter-graphene wall spacing is from our reference *11b*.

**Figure 5.** Schematic representation of a solar thermal and photovoltaic field to drive both water purification and C2CNT splitting of $CO_2$ to useful products.

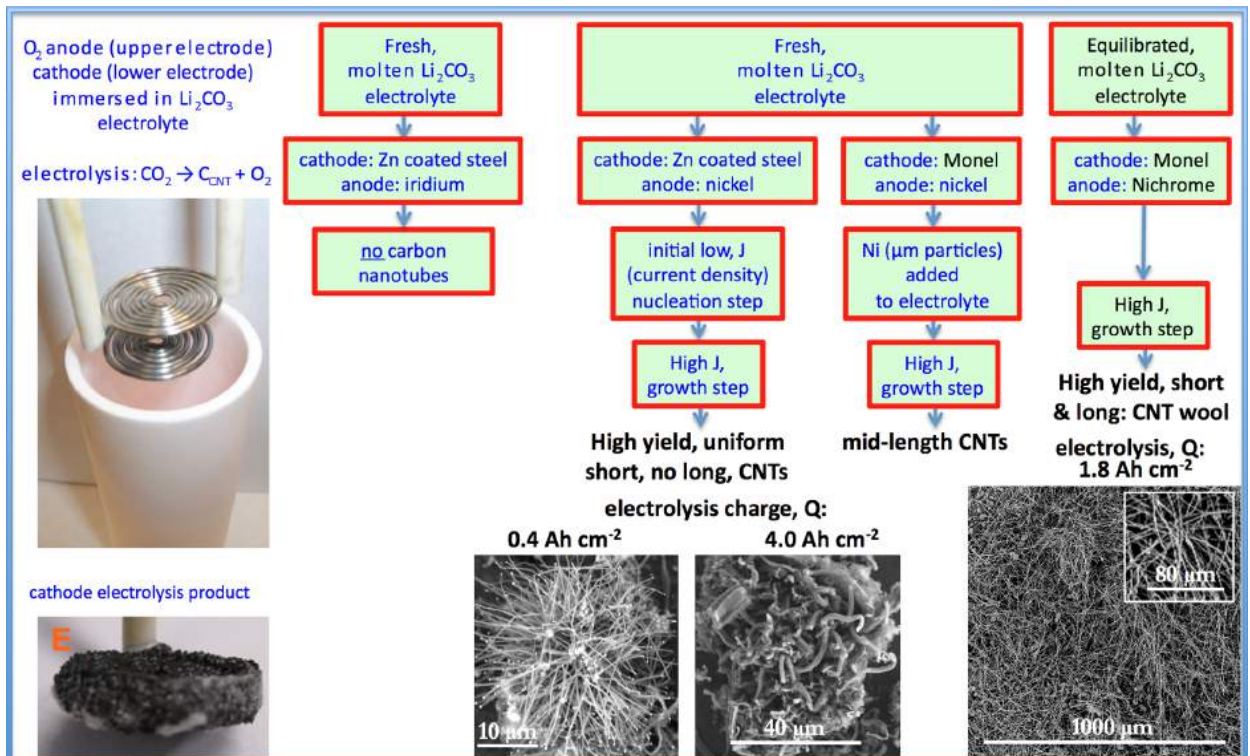

Figure 1

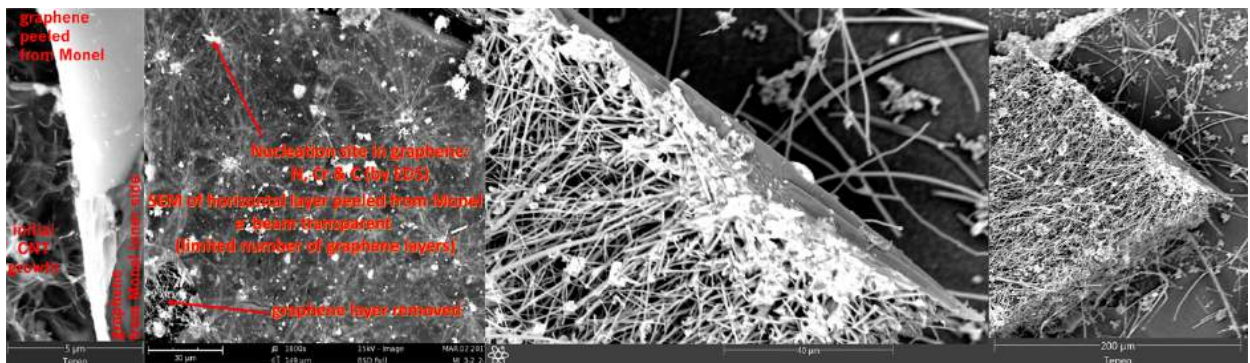

Figure 2

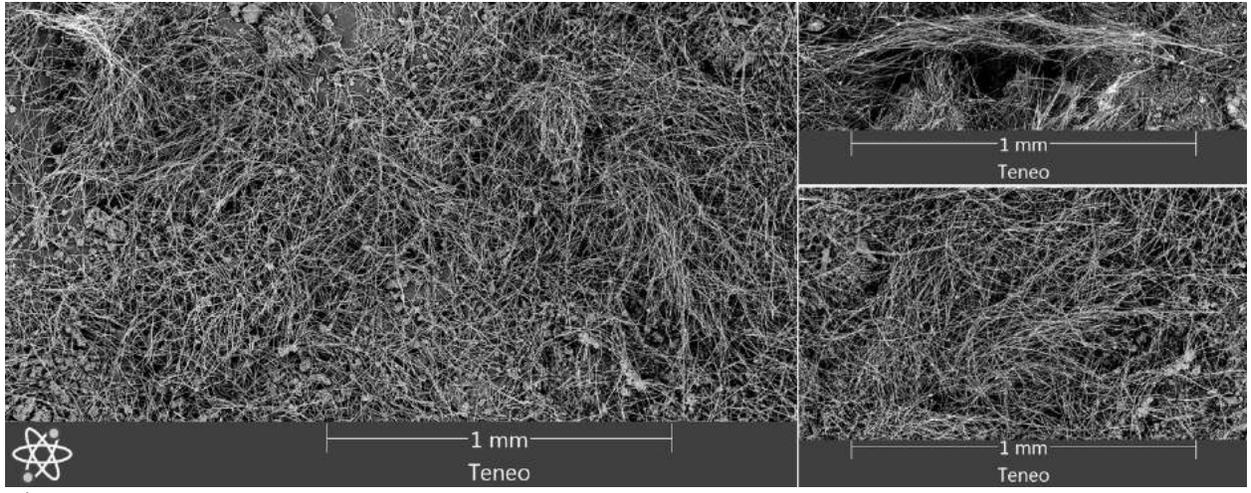
Figure 3

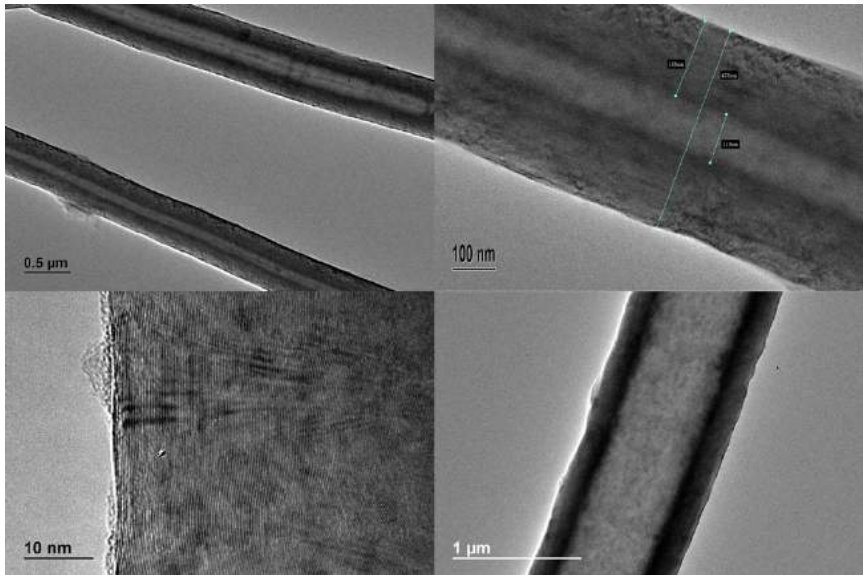
Figure 4

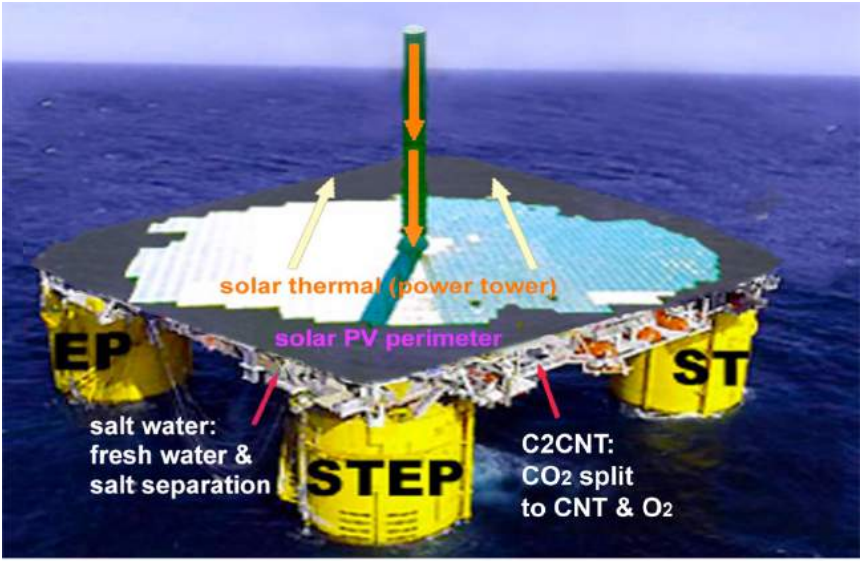

Figure 5

# Supplemental Information accompanying:

## Carbon Nanotube Wools Directly from $CO_2$ By Molten Electrolysis: Value Driven Pathways to Carbon Dioxide Greenhouse Gas Mitigation

M. Johnson[1], J. Ren[1], M. Lefler[1], G. Licht[1], J. Vicini[1], S. Licht[1],*(slicht@gwu.edu)
[1]Dept. of Chemistry, George Washington University, Washington DC 20052.

### I. C2CNT electrosynthesis with bare (uncoated) cathodes and without pre-electrolysis low current activation:

In the prior electrosynthetic methodology, the cathode consists of a galvanized (zinc coated) steel electrode, and an initial low current (0.05 A for 10 minutes, 0.1A for 10 minutes, 0.2A for 5 minutes, and 0.4A for 5 minutes) series of steps is applied to grow Ni nucleation sites on the cathode, followed by a longer, constant current (controlled at 0.1 to 0.2 A cm$^{-2}$). As one example of the prior methodology the electrolysis is conducted with a lithium metaborate additive to the electrolyte, that is to the 50 g of $Li_2CO_3$, either 1.5 g, 3 g, 5g, 8 g, or 10g of $LiBO_2$ is added to the electrolyte. The SEM observed morphology of the products remains unchanged with these various levels of $LiBO_2$ addition, and consists of 5 to 50 µm long carbon nanotubes (CNTs) as exemplified on the left of Figure S1. However, as we have recently), the boron addition to the electrolyte boron-dopes the CNTs (as determined by a Raman peak shift in the G band *(11d)*) seen in the top right of Figure S1) and increases their electrical conductivity by a factor of ten as summarized in the bottom right of Figure S1.

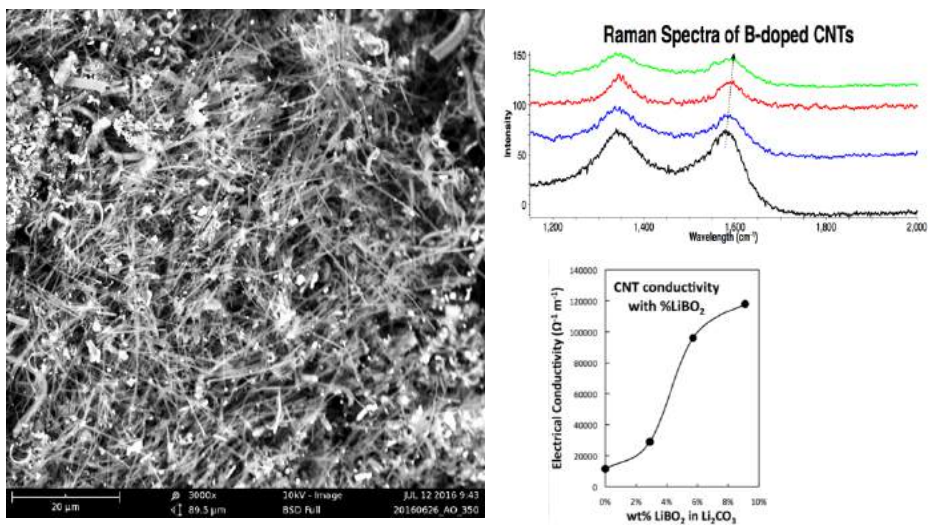

**Figure S1.** Properties of boron doped CNTs formed by electrosynthesis in molten carbonate. SEM (top) and Raman spectra (left) of B-doped CNTs formed by 1 Ah electrolysis at a 5 cm$^2$ cathode in 5g $LiBO_2$ and 50 g $Li_2CO_3$ at 770°C. Right: the electrical conductivity of CNTs grown with an increasing concentration of $LiBO_2$ dissolved in the $Li_2CO_3$ electrolyte. Note that we had previously reported the anode and cathode surface area each as 10 cm$^2$. More specifically this was the total (two sided) exposed surface area, whereas the surface area facing each electrode is 5 cm$^2$.

In this study several combined C2CNT effects (elimination of the cathode zinc coating and pre-electrolysis activation steps, cathode substrate composition, electrolyte aging, and choice of anode substrage composition) lead to the first production of CNT wool. In this supplementary section, the first several of these effects effects (removal of the cathode zinc coating effect, elimination of the pre-electrolysis low-current activation steps and cathode substrate composition choice). As a first step towards the new synthesis electrolyses were conducted without zinc coated cathodes, which had been considered as a necessary component to the synthesis (Dey, et al., 2016; Ren, et al., 2015; Ren, et al., 2017; Ren and Licht, 2016; Wu, 2016, Licht, et al., 2016), but instead with fine (3-5 µm) Ni metal powder added directly as additional transition metal to the electrolyte to compensate for the lack of the zinc activating agent. Less than 1% by mass added Ni was sufficient (0.1 to 10% Ni addition was explored, and larger than 5 µm added Ni powder was less effective than the 2-5 µm powder) to promote CNT growth. Removal of the zinc coating constraint opened the pathway to explore other (than steel) metals as cathode substrates and in particular certain metal substrates promoted a longer CNT product. The previous galvanized steel cathode had a zinc coating and as zinc has a 420°C melting point which is less than the temperature of the molten electrolyte ($\geq$ 750°C) in which the cathode is immersed, liquid zinc could form. In that case the liquid zinc could leave the electrode and helps initiate the dissolution of nickel from the anode or formation of carbon (Ren, et al., 2015; Ren, et al., 2017; Ren and Licht, 2016; Wu, 2016). In lieu of the zinc, the direct, addition of Ni powder to the electrolyte provides sharper control of the initiation of CNT growth than the previous methodology which utilized the release of Ni during the initial gradual formation of a stable Ni oxide layer at the anode which had been observed to require a gradual increase of electrolysis current to yield a high formation of the CNT product at the cathode. Here the higher steady-state electrolysis current can be initially and continuously applied without the need for that lower current density activation of CNT growth.

As seen in Figure S2 by SEM of the cathode product, even a low level (0.1 wt%) of the Ni powder added to the $Li_2CO_3$ electrolyte promotes CNT. To ensure that no Ni is in the system other than that added as Ni powder, an iridium anode, rather than Ni or Ni alloy anode, was used in this electrosynthesis, and we've previously noted that Ir is also an effective (albeit expensive) oxygen electrode for the carbonate system. As seen in the top panel of the figure, without the Ni powder (and without the zinc cathode coating) no CNT product is observed. However, with the added Ni powder (and still without the zinc cathode coating), in the middle amd lower panels, it seen these product CNTs are highly uniform and of high purity CNTs.

Each of the subsequent experiments in this supplementary section are conducted on various cathodes, each by electrolysis with 0.4 wt% of 3-5 µm Ni powder utilized in the 770°C molten lithium carbonate and in each case the electrolyses were conducted at constant current (without an initiating series of activation increasing constant current steps). As seen in Figure S3, while Monel and copper both produce a high yield of CNTs, the CNT morphology is entirely different after 1.5 hours of electrolysis time (at 1-amp constant current between the 5 cm$^2$ electrodes). The copper cathode forms thin, tangled CNTs, while the Monel cathode forms uniform, thicker and straight CNTs. A pure Ni cathode (not shown) produces a result similar to that of copper although the CNT yield (80 to 85%) is less than that of the $\geq$ 85% yield of the copper cathode. As shown on the left side of Figure S4, a Nichrome cathode provides straighter CNTs than a pure nickel cathode. The nickel chromium cathode continues to produce straight

CNTs during intermediate duration electrolyses, but unlike the Monel cathode CNTs from a nickel chromium substrate cathode did not continue to grow during extended electrolytes. Iron oxide (unlike nickel oxide) is highly soluble in molten carbonate and we have previously shown that its addition to the electrolyte results, in an <u>uncontrolled</u> growth of a profusion of nanostructures (Ren, et al., 2015; Ren, et al., 2017; Ren and Licht, 2016; Wu, 2016). A pure iron cathode substrate results in a similar product as seen in the right side of Figure S4. Figure S5 presents the product generated at several different cathodes subsequent to extended (12 Ah) electrolyses. As seen a titanium cathode yielded shorter and only moderate quality CNTs, while a graphite foil cathode yielded high quality, but shorter (than a Monel Cathode) CNTs subsequent to these extended electrolysis.

Of the eight cathodes examined here (Monel, steel, iron, nickel, nickel chromium, copper, titanium, and graphite) only Monel exhibited a CNT product whose length increases (and was approximately linear) with the increase in integrated electrolysis charge density.

Subsequent to this screening of the electrolysis cathode substrate, two further significant advances to the C2CNT process were found as included in the main text of this study. 1) Uncoated (bare, without zinc) substrate cathodes can generate a CNT product <u>without any</u> Ni powder added to the electrolyte, when the molten electrolyte is "aged" (left molten prior to electrolysis) for 24 hours prior to electrolysis (longer periods did not further improve CNT quality). Presumably in this case the low level of nickel dissolving from the anode is sufficient to migrate and act as nucleation at the cathode, and this originates while the anodic nickel oxide layer is established at the start of the electrolysis. 2) The quality (length and quantity of CNT) improves when the anode consists of Nichrome, rather than a pure nickel anode, and both Ni and Cr are then observed by EDS at the cathode CNT nucleation sites.

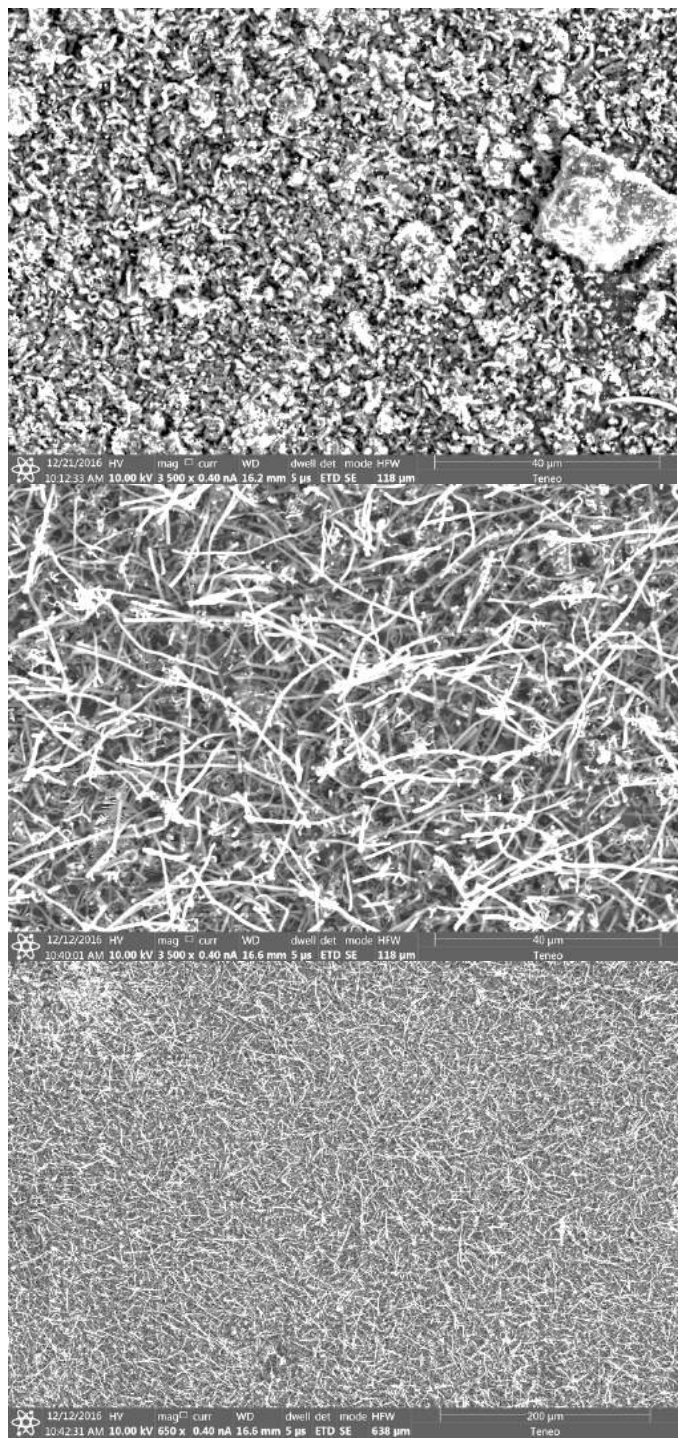

**Figure S2**. Top: without added Ni powder. a bare (zinc free) cathode does not form an observable CNT product from a fresh molten $Li_2CO_3$ electrolyte, but with appropriate choice of substrate can form a highly uniform CNT product (middle and lower panel) with the addition of a low level (0.1 wt%) 3-5 µm Ni powder to the electrolyte. 1.2 Ah cm$^{-2}$ electrolyses are conducted using an Ir anode (rather than Ni anode, to ensure the anode does not introduce nickel to the electrolyte) and a Monel cathode in 770°C $Li_2CO_3$.

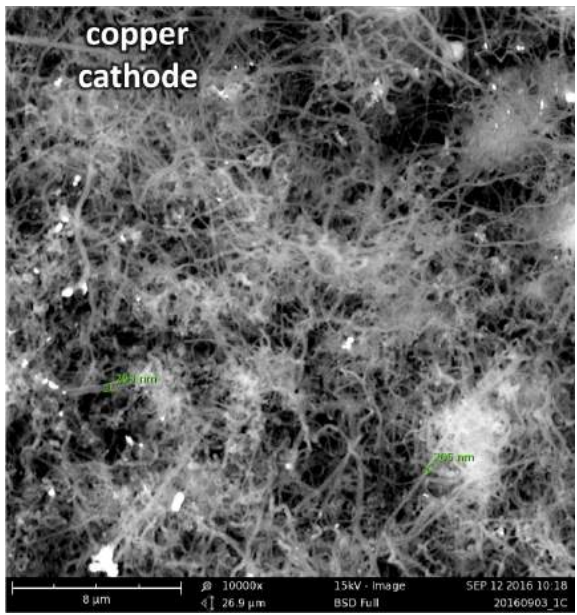 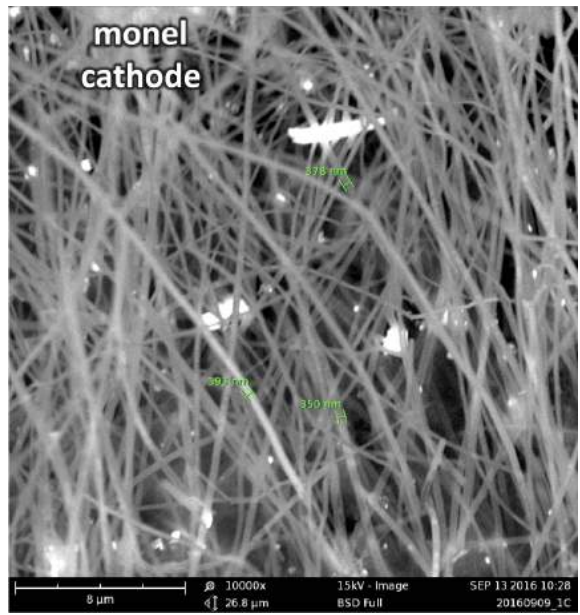

**Figure S3.** Comparison of the CNT product formed respectively at a copper (left side) and Monel (right side) cathode during short duration 0.3 Ah cm$^{-2}$ electrolysis in 770°C Li$_2$CO$_3$.

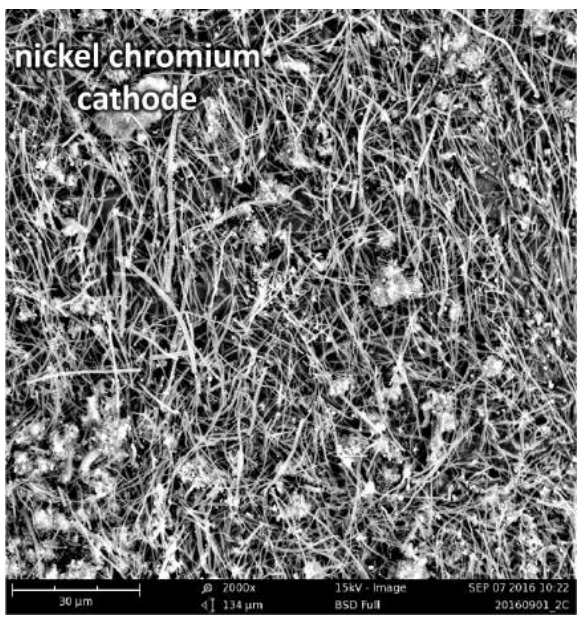 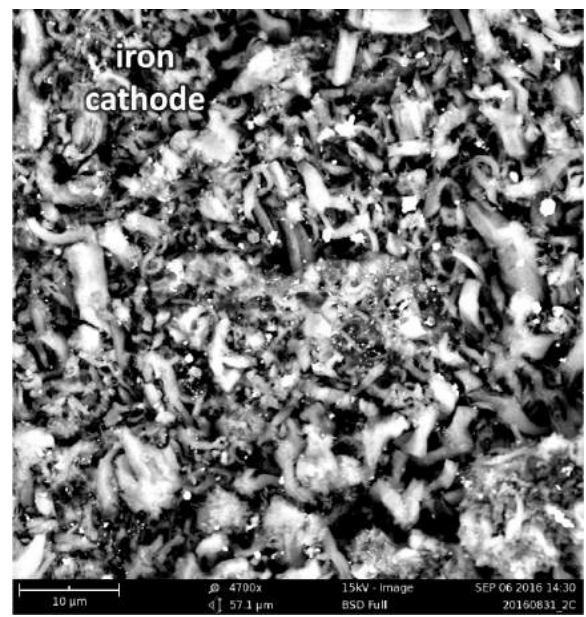

**Figure S4.** Comparison of the CNT product formed respectively at a nickel chromium alloy (left side) and iron (right side) cathode electrolyses in 770°C Li$_2$CO$_3$. The product on Nichrome is formed during intermediate duration (0.8 Ah cm$^{-2}$), while at an iron cathode, as shown even for short duration electrolysis (0.2 Ah cm$^{-2}$), the carbon product is highly heterogeneous when formed.

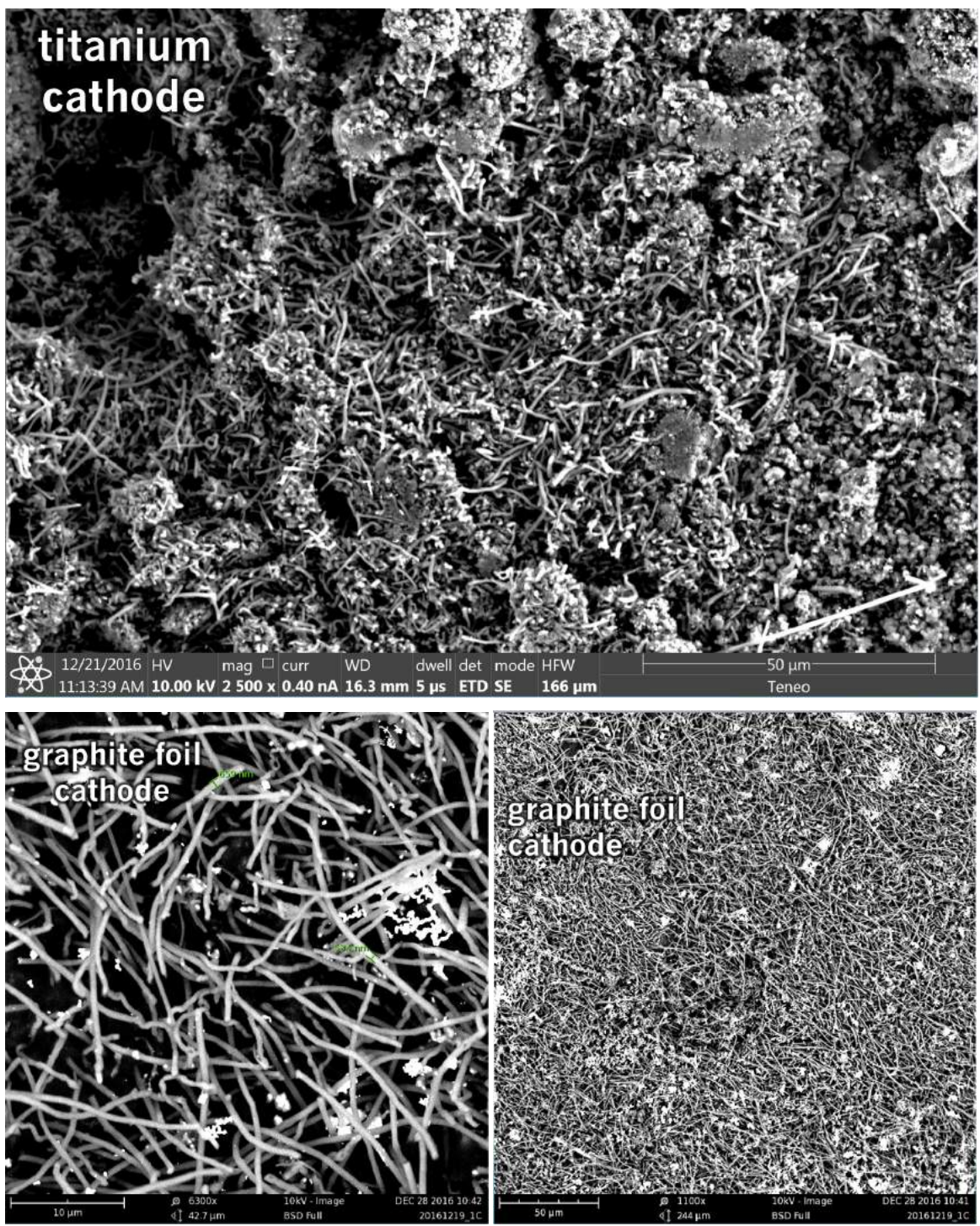

**Figure S5.** Comparison of the CNT product formed respectively at a titanium (left side) and graphite foil (right side) cathode extended (2.4 Ah cm$^{-2}$) electrolyses in 770°C Li$_2$CO$_3$. The graphite foil is cut as a 5 cm$^2$ disc, while the titanium (and copper, Monel, iron, steel, nickel or nickel chromium cathode substrate) is coiled wire 5 cm$^2$ discs.

The physical chemical environment of the conventional CVD CNT synthesis is different than that of the new C2CNT synthesis in most aspects. The latter is an electrochemical process, while the former is chemical. The latter utilizes $CO_2$, while the former utilizes organics as the reactant, and the latter occurs at the liquid/solid interface, while the latter generally occurs at a gas/solid interface. There are also significant subtle differences. C2CNT provides a higher density of reactive carbon (the molten carbonate electrolyte) near the growth interface, and while an electric field may, or may not, be applied to the substrate during CVD CNT growth, there is always an intense electric field rapidly decreasing through the double layer adjacent to the cathode during C2CNT growth. Despite these differences, it is fascinating that several phenomena observed here that promote C2CNT processes also appear to have a similar affect on CVD CNT growth processes, and other CVD advances suggest pathways for further C2CNT improvements. Few-layered graphene/multiwalled CNT structures have been observed to form by CVD metal (Ni) substrates (Wang, et al., 2013). In CVD, larger multiwall CNT (more than ten walled MWCNTs) almost exhibit tip growth with the nucleating metal in the lead, rather than base growth as it is thought be that larger nano-particles adhere less to substrate), while few walled MWCNTs ($\leq$ 7) or less almost always form by base growth (Ren, 2015). In CVD, rapid pre-heating of the catalyst before introduction of the feed gas resulted in the growth of longer/higher yield CNTs, and applied electric field helped promote the growth of more aligned and straighter CNTs (Huang, et al., 2003). Metal nano-particles can grow and get become trapped or move along in CNT, consuming and requiring more catalyst or stopping the growth process (Gozzi, et al., 2006). In single walled CNT CVD growth studies, a nucleation period of 5-10 seconds occurs at the start of the CNT growth. Initially, this nucleation period requires more carbon in proximity to the growth region and less during the next period consisting of a simultaneous CNT growth and repair stage (Qi, et al., 2007). The growth of longer CNTs was facilitated by the initial preparation of micrometer catalyst islands on a substrate CNT (Kong, et al., 1998). Bi-metallic catalysts promote CNT growth better than single metals alone. Splitting the catalyst into two groups, ones that help nucleation more and one that helps growth and repair more leads to the best two metals (choose the best from each group) and led to the following organization (Wei-Qiao Deng, et al., 2004):

1. (Order of) best: Ni+Co > Ni + Pt >> Cu + Co
   Ni+Co > Ni + Fe  ~ Ni >> Fe
   Ni + Co > Ni + Pt > Ni + Cu
2. Nucleation: Co > Pt > Ni slightly less > Cu (poisons)
   Theory: Mo > Cr > Co > Pt > Re > Fe > Ni > Pd
3. Growth & repair: Ni > Co > Pt> Cu
   Theory: Ni + Mo > Ni + Cr > Ni + Co > Ni+ Pt > Ni + Rd > Ni +Fe > Ni > Fe
   & Fe + Mo > Fe + Cr > Fe + Co > Fe + Pt > Fe +Rh > Fe
   & Ni+Mo > Fe + Mo > Co+Mo>Co (2-4 by theory, later Co+Mo > Co

Due to a lower melting point, copper can promote CVD CNT growth under certain conditions (Zhou, et al., 2006). Different metals lead to different diameter CNTs (Zhang, et al., 2008). Islands were found to form on the substrate and their size can lead to optimization (Hofmann, et al., 2007).

## II. C2CNT Intermediate length CNTs with intermediate integrated electrolysis charge transfer:

Figure 6S presents an intermediate stage of the C2CNT synthesis advancement resulting in improved CNT yield and improved CNT length. In this intermediate advancement of synthesis, a bare Monel (rather than galvanized steel) substrate was used, a 2-step (0.05A/13 min, 0.25A/12min), rather than the original 4-step electrolysis pre-activation was utilized, 0.4g (0.8 wt%) of Ni powder was added to the 50g of 770°C $Li_2CO_3$. electrolyte, and an Ir anode, and a somewhat higher current density and integrated charge was used was used (1A through the 5 cm$^2$ electrode (0.2 A cm$^{-2}$), for 6 hours (integrated charge of 1.2 Ah cm$^{-2}$). The higher constant current resulted in a higher (1.6V) potential during the electrolysis. Up to 10 wt% added $LiBO_2$ was observed to have no effect on the observed CNT morphology, but as described in section 1 enhances their electrical conductivity. The electrolysis product shown in Figure 6S contains 3g $LiBO_2$ added to the electrolyte. As seen in the figure, even without the additional improvements of mixed Ni/Cr, rather than just Ni, nucleation, and without electrolyte aging, the CNT quality is high and the tangled CNTs are longer, ranging from 20 to over 200 μm in length.

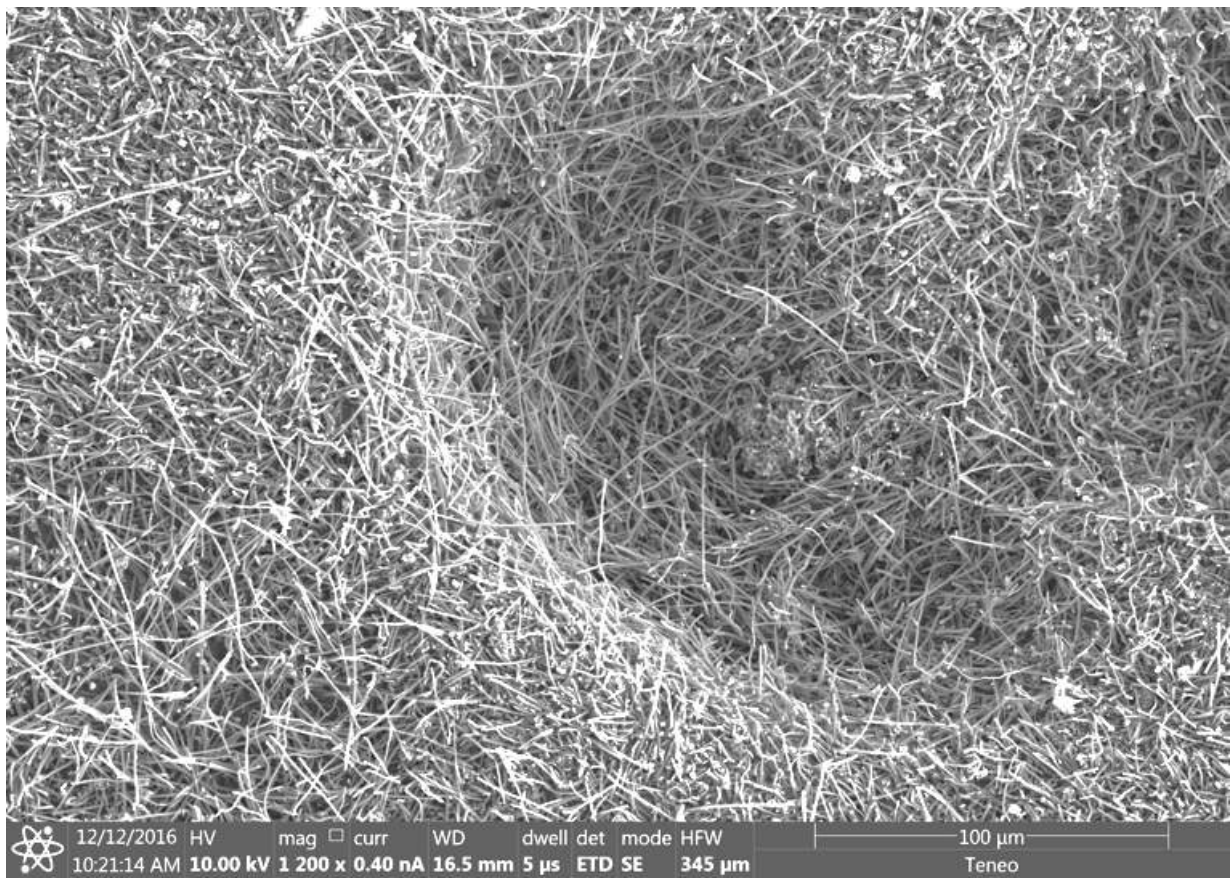

**Figure S6.** SEM of B-doped CNTs formed by 6 Ah electrolysis at a 5 cm$^2$ Monel cathode in 5g $LiBO_2$ and 50 g $Li_2CO_3$ at 770°C.

### III. Improved admixing of sulfur, nitrogen and phosphorous (in addition to boron) to carbon nanotubes:

As previously reported (Ren, et al., 2017), unlike in pure carbonate, <u>no</u> carbon product (CNT or otherwise) was observed to form at the cathode during the electrolysis of 1 mol % (or 3, or 5 mol %) $Li_2SO_4$ in 770°C $Li_2CO_3$. The observed potentials at 1 A are lower with higher [$Li_2SO_4$] (and are lower than the 1-2volt electrolysis potential observed without $Li_2SO_4$). This lack of CNT formation is in accord with the electronegativity of sulfur compared to carbon, which favors the thermodynamic formation of the former compared to the latter. Here, we report the first successful electrosynthesis of sulfur heteroatom CNTs. To improve the energetics of carbon formation, the concentration of sulfate is decreased (relative to carbonate), and this creates a pathway to the observed formation of sulfur containing CNTs. Specifically, the Figure S7 left side presents sulfur containing CNTs from molten carbonate electrolysis with 0.1 mole% sulfate subsequent to a 2-hour electrolysis at 1 A (using the conventional galvanized steel cathode and Ni 200 wire anode and without added Ni metal powder). EDS of the CNT product measured 0.1 mole % of sulfur in the CNT product. As in previous experiments, prior to this higher current extended electrolysis, cathode nucleation was facilitated by an application of lower constant currents sequentially applied (each for 10 minutes) and increased from 0.05, 0.10, 0.25 to 0.5 A. As previously observed with successful (non-sulfur containing electrolyte) CNT electrolyses (Ren, et al., 2015; Ren, et al., 2017; Ren and Licht, 2016; Wu, 2016), the initial 10-minutes lowest current electrolysis occurred at a potential of 0.4 to 0.5 V, which is consistent with the expected nucleation by Ni on the cathode while each of the subsequent increasing constant currents occurred at increasing potentials between 1 to 2 V.

We improve on our recently observed (Ren, et al., 2017) electrolytic formation of P-heteroatom CNTs from lithium metaphosphate dissolved in a lithium carbonate electrolyte and report the first evidence of phosphorous in the CNT product. The use of $LiPO_3$ is observed to facilitate salt dissolution in the lithium carbonate electrolyte. Variations which led to the improved length and yield of P-containing CNTs include an increase from the previous 1 % to 5 mol % if $LiPO_3$, and the use of a Monel, rather than galvanized steel cathode. On the right side of Figure S7, the product P-heteratom long (300-600 µm) are produced with intermediate 0.8 Ah cm$^{-2}$ charge at a low current density of 0.03 A cm$^{-2}$; a conventional (Ni 200) anode and <u>no</u> Ni powder was added to the electrolyte during this C2CNT synthesis, EDS of the CNT product measured 0.3 mole % of phosphorous in the CNT product. This is substantially lower than the electrolytic concentration of phosphorous, and the P-heteroatom may provide a poor lattice match to the CNT.

A CNT product is also observed from electrolysis of $LiNO_3$ in the 770°C $Li_2CO_3$ electrolyte. In this case, the yield of CNTs improves with a 5 mole %, compared to a 1 mole %, dissolution of $LiNO_3$ within the electrolyte. EDS analysis of nitrogen within the product is indicated, but not confirmed, as the EDS instrumentation was broad and not capable or resolving the near lying carbon (12.01) and nitrogen (14.01) peaks. SEM and an extended analysis of the nitrogen CNT product, and also a detailed analysis of the doped CNT growth mechanism, further elemental probes of each of the heteroatom modified CNTs, and applications of these CNTs will each be presented in an expanded, future study.

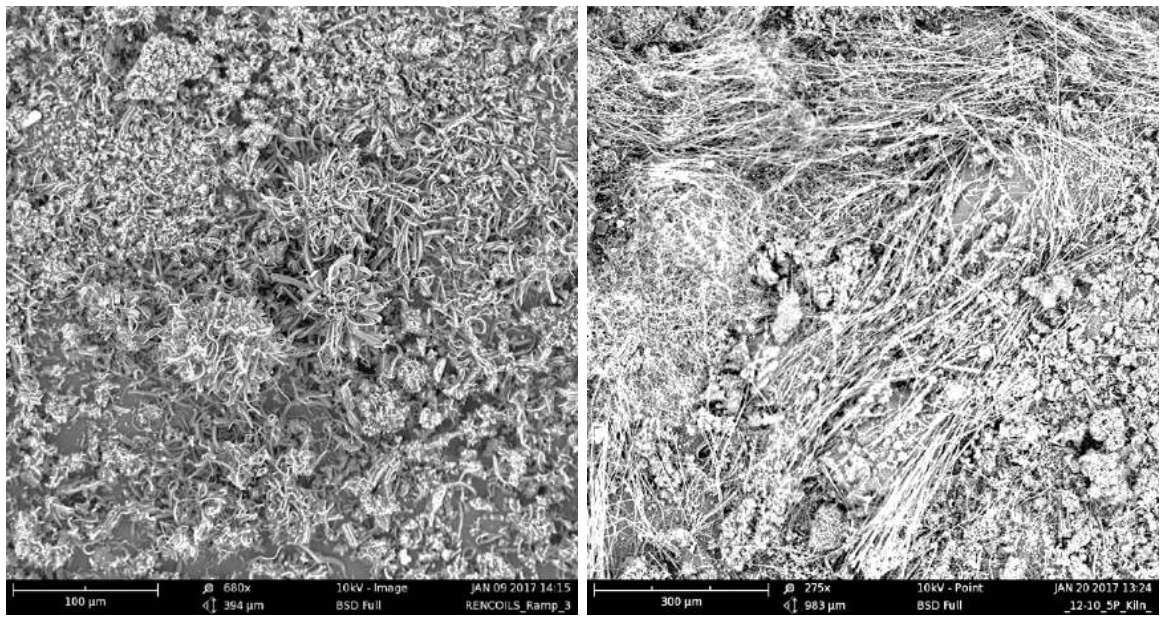

**Figure S7.** Left: SEM of the S-heteroatom product formed by 0.4 Ah cm$^{-2}$ electrolyis at a galvanized steel cathode in 50g of 770°C $Li_2CO_3$ containing 0.074g $Li_2SO_4$. Right: SEM of the P-heteroatom product formed by 0.8 Ah cm$^{-2}$ electrolyis at a Monel cathode in 50g of 770°C $Li_2CO_3$ containing 0. 5 mol % $Li_2PO_4$.

The successful electrosynthesis of CNTs containing the heteroatoms of sulfur, phosphorus or boron, and likely nitrogen, is observed. Doped CNTs can have unusual, useful properties including high electrical conductivity, catalysis, heavy metal removal, enhanced oxygen kinetics and improved charge storage.

## IV. Scalability of the C2CNT process:

As an electrochemical process, C2CNT is linearly scale-able with our increasing area of the electrolysis chamber (Licht, 2017), Figure S8.

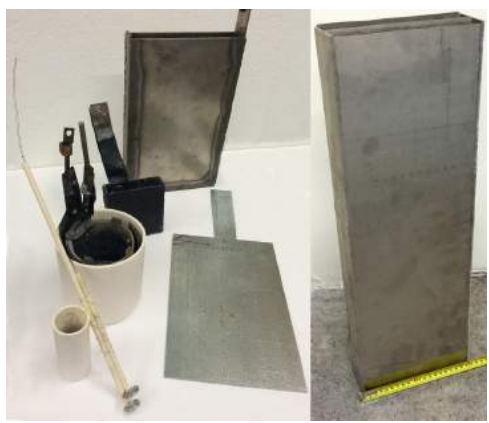

**Figure S8.** The evolution of the electrolysis chamber. Earlier versions can be seen in the front on the left, and later versions in the back and to the right. The rectangular electrolysis chambers use the interior walls as the air electrode (Licht, 2017).

The thermodynamic and cost savings of new and retrofit cement, gas and coal power plants has been analyzed (Lau, et al., 2016; Licht, 2017). Industrial plant retrofit provides a ready source of hot $CO_2$ for electrolysis, the oxy-fuel energy benefits of the C2CNT co-product $O_2$ looped back into the plant, the value of the CNT product, and further impetus for intermediate C2CNT scale-up as illustrated if Figure S9.

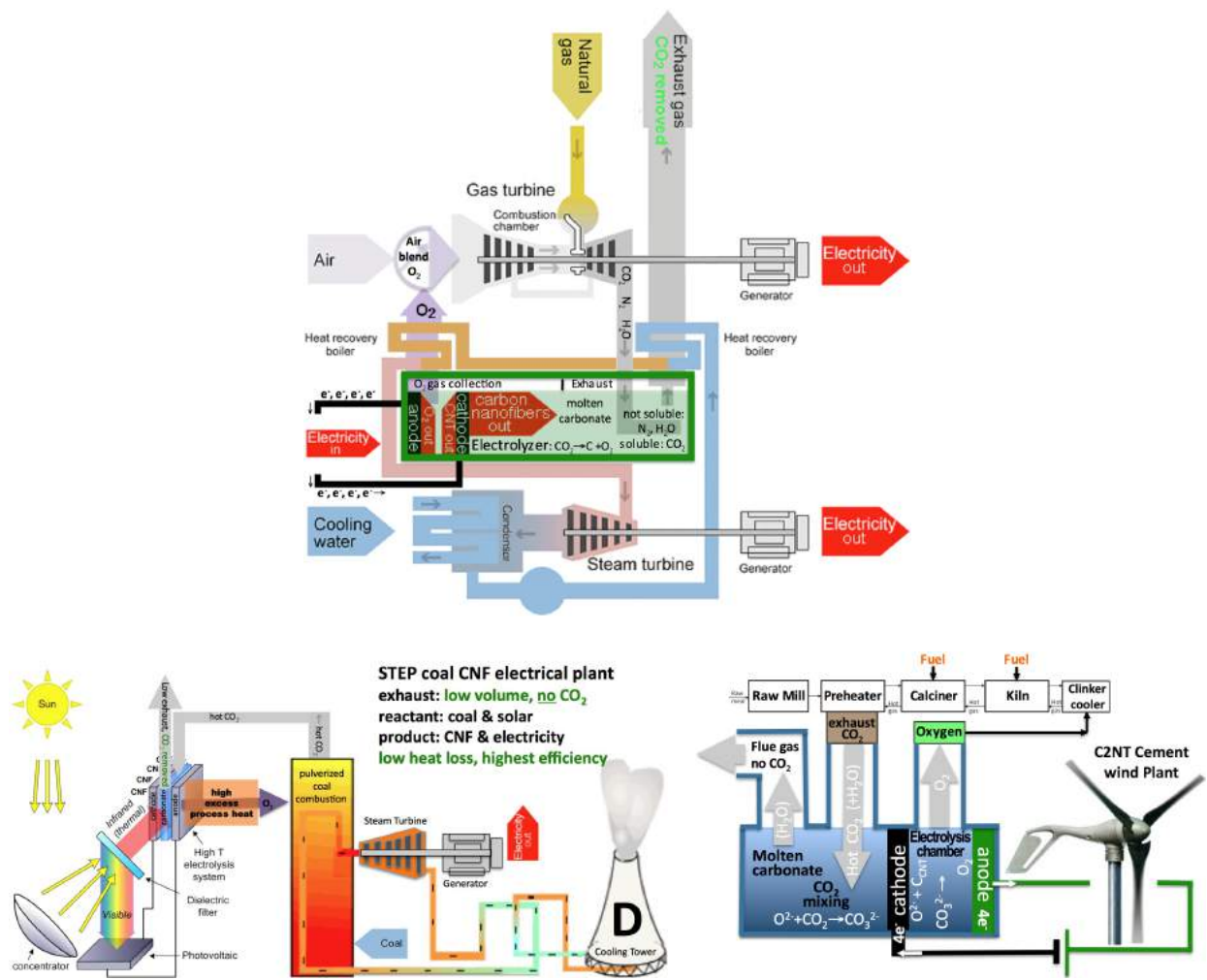

**Figure S9.** Top: Schematic of a CNT combined cycle power plant (Lau, et al., 2016). Middle: Transforming $CO_2$ emissions from a coal combustion plant into CNTs using solar energy (Lau, et al., 2016). Bottom: C2CNT Cement wind plant: The full oxy-fuel configuration is shown. The plant does not emit $CO_2$, and over time cement produced absorbs $CO_2$. Hence the process is carbon negative, which compares favorably to the large positive carbon signature of conventional cement plants (Licht, 2017).

C2CNT atmospheric mitigation does not require pre-concentration of the CO₂. Heat is then provided using photovoltaic discarded thermal sunlight (Li, et al., 2011) as illustrated if Figure S10.

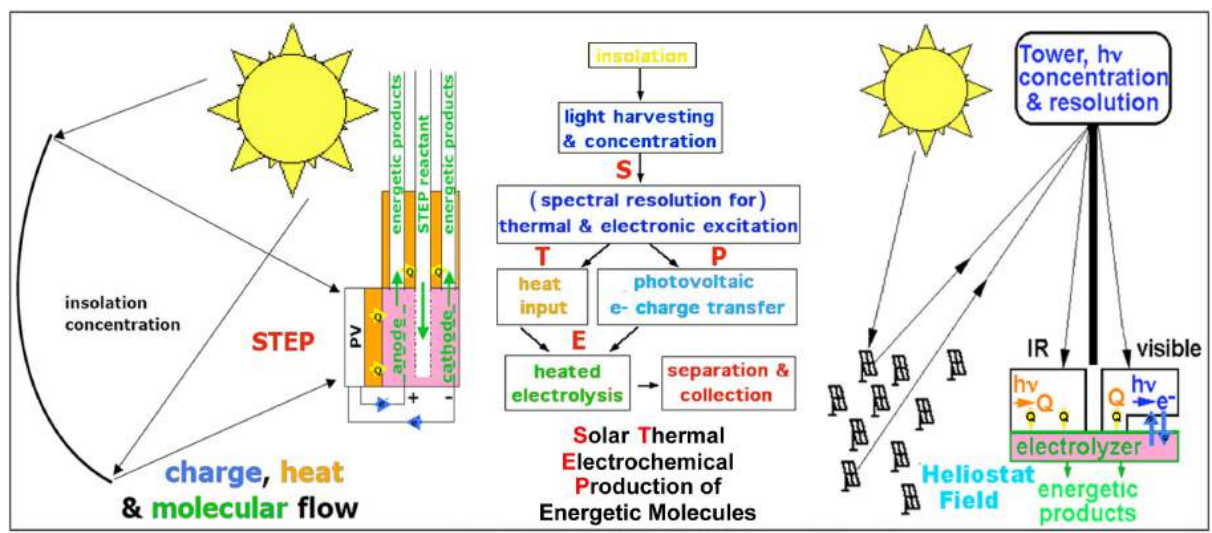

**Figure S10.** Global use of sunlight to drive the formation of energy rich molecules. Left: Charge, & heat flow in STEP: heat flow (yellow arrows), electron flow (blue), and reagent flow (green). Right: Beam splitters redirect sub-bandgap sunlight away from the PV onto the electrolyzer (Licht, et al., 2011).

The following section is expanded by addition of a CO₂ availability concluding paragraph from "Scalability of STEP Processes" section of the reference Licht, 2011, which in turn was expanded from the text and Supporting Information from the reference Licht, et al., 2010. Note, since this 2011 analysis, subsequent total estimates of the extent of CO₂ released in industrial revolution have increased to over 1.1 teratons.

STEP can be used to remove and convert CO₂. As with water splitting, the electrolysis potential required for CO₂ to CO splitting falls rapidly with increasing temperature, and we have shown that a photovoltaic, converting solar to electronic energy at 37% efficiency and 2.7V, may be used to drive three CO₂ splitting, lithium carbonate electrolysis cells, each operating at 0.9V, and each generating a 2 electron CO product. The energy of the CO product is 1.3V (eq 1), even though generated by electrolysis at only 0.9V due to synergistic use of solar thermal energy. At lower temperature (770°C, rather than 950°C), carbon, rather than CO, is the preferred product, and this 4 electron reduction approaches 100% Faradaic efficiency.

The CO₂ STEP process consists of solar driven and solar thermal assisted CO₂ electrolysis. Industrial environments provide opportunities to further enhance efficiencies; for example fossil-fueled burner exhaust provides a source of relatively concentrated, hot CO₂. The product carbon may be stored or used. STEP represents a new solar energy conversion processes to produce energetic molecules. Individual components used in the process are rapidly maturing technologies including wind electric, molten carbonate fuel cells, and solar thermal technologies.

It is of interest whether material resources are sufficient to expand the process to substantially impact (decrease) atmospheric levels of $CO_2$. The buildup of atmospheric $CO_2$ levels from a 280 to 392 ppm occurring over the industrial revolution comprises an increase of $1.9 \times 10^{16}$ mole ($8.2 \times 10^{11}$ metric tons) of $CO_2$, and will take a comparable effort to remove. It would be preferable if this effort results in useable, rather than sequestered, resources. We calculate below a scaled up STEP capture process can remove and convert all excess atmospheric $CO_2$ to carbon.

In STEP, 6 kWh m$^{-2}$ of sunlight per day, at 500 suns on 1 m$^2$ of 38% efficient CPV, will generate 420 kAh at 2.7 V to drive three series connected molten carbonate electrolysis cells to CO, or two series connected series connected molten carbonate electrolysis cells to form solid carbon. This will capture $7.8 \times 10^3$ moles of $CO_2$ day$^{-1}$ to form solid carbon (based on 420 kAh · 2 series cells / 4 Faraday mol$^{-1}$ $CO_2$). The $CO_2$ consumed per day is three fold higher to form the CO product (based on 3 series cells and 2 F mol$^{-1}$ $CO_2$) in lieu of solid carbon. The material resources to decrease atmospheric $CO_2$ concentrations with STEP carbon capture, appear to be reasonable. From the daily conversion rate of $7.8 \times 10^3$ moles of $CO_2$ per square meter of CPV, the capture process, scaled to 700 km$^2$ of CPV operating for 10 years can remove and convert all the increase of $1.9 \times 10^{16}$ mole of atmospheric $CO_2$ to solid carbon. A larger current density at the electrolysis electrodes, will increase the required voltage and would increase the required area of CPVs. While the STEP product (chemicals, rather than electricity) is different than contemporary concentrated solar power (CSP) systems, components including a tracker for effective solar concentration are similar (although an electrochemical reactor, replaces the mechanical turbine). A variety of CSP installations, which include molten salt heat storage, are being commercialized, and costs are decreasing. STEP provides higher solar energy conversion efficiencies than CSP, and secondary losses can be lower (for example, there are no grid-related transmission losses). Contemporary concentrators, such as based on plastic Fresnel or flat mirror technologies, are relatively inexpensive, but may become a growing fraction of cost as concentration increases. A greater degree of solar concentration, for example 2000 suns, rather than 500 suns, will proportionally decrease the quantity of required CPV to 175 km$^2$, while the concentrator area will remain the same at 350,000 km$^2$, equivalent to 4% of the area of the Sahara Desert (which averages ~6 kWh m$^{-2}$ of sunlight per day), to remove anthropogenic $CO_2$ in ten years.

A related resource question is whether there is sufficient lithium carbonate, as an electrolyte of choice for the STEP carbon capture process, to decrease atmospheric levels of $CO_2$. 700 km$^2$ of CPV plant will generate $5 \times 10^{13}$ A of electrolysis current, and require ~2 million metric tonnes of lithium carbonate, as calculated from a 2 kg/liter density of lithium carbonate, and assuming that improved, rather than flat, morphology electrodes will operate at 5 A/cm$^2$ (1,000 km$^2$) in a cell of 1 mm thick. Thicker, or lower current density, cells will require proportionally more lithium carbonate. Fifty, rather than ten, years to return the atmosphere to pre-industrial $CO_2$ levels will require proportionally less lithium carbonate. These values are viable within the current production of lithium carbonate. Lithium carbonate availability as a global resource has been under recent scrutiny to meet the growing lithium battery market. It has been estimated that the current global annual production of 0.13 million metric tons of LCE (lithium carbonate equivalents) will increase to 0.24 million tons by 2015. Sodium carbonate is substantially more available, and as noted can be combined with lithium carbonate for molten $CO_2$ splitting. Low velocity natural wind speeds are sufficient to move this air to C2CNT processors. A 100 km by 100 km area with wind moving through it at 2 km per hour will deliver over a teraton of $CO_2$ during a decade.